\documentclass[fleqn,twoside]{article}
\usepackage{amssymb}
\usepackage{graphicx}
\usepackage{espcrc2}

\graphicspath{{plots/}}
\DeclareGraphicsExtensions{.eps.gz,.eps,.ps,.ps.gz}

\newcommand{\lwig}{\mbox{\,\raisebox{.3ex}
    {$<$}$\!\!\!\!\!$\raisebox{-.9ex}{$\sim$}\,}}
\newcommand{\gwig}{\mbox{\,\raisebox{.3ex}
    {$>$}$\!\!\!\!\!$\raisebox{-.9ex}{$\sim$}}\,}

\newif\ifhepph

\hepphtrue

\begin{document}

\title{
\ifhepph\vbox to 0pt{{\vss\flushright\normalsize\rm
DESY 02-158\\
hep-ph/0210209\\
\endflushright}}\fi
Vacuum structure and high-energy scattering\ifhepph\thanks{Talk presented at QCD 02, Montpellier/F, July 2002.}\fi}

\author{A.~Ringwald
\address{Deutsches Elektronen-Synchrotron DESY, 
Notkestra\ss e 85, D--22607 Hamburg, Germany}}

\begin{abstract}
This short review deals with the manifestations of the vacuum structure of non-Abelian gauge theories 
in high-energy scattering. Specifically, it concentrates on instanton-induced hard scattering processes,  
both in the electroweak gauge theory and in QCD. Soft scattering processes in QCD and their connection to 
models of semi-hard vacuum fluctuations  are also briefly discussed.    
\end{abstract}

\maketitle

\section{INTRODUCTION}

Non-Abelian gauge theories like QCD are known to possess a rich 
vacuum structure. Notably, there are topologically 
non-trivial vacuum fluctuations of the gauge fields, whose simplest
examples are instantons~\cite{Belavin:fg}.
Instantons have been argued to play an important role in various
long-distance aspects of QCD, such as giving a possible solution to 
the axial $U(1)$ problem~\cite{'tHooft:fv} or being at work in
$SU(n_f)$ chiral symmetry breaking~\cite{Shuryak:1981ff
}.  
Recently, also a number of QCD lattice studies elucidated 
the topological structure and 
the instanton content of the vacuum~\cite{Teper:1999wp}.

Another prominent feature of the QCD vacuum is its apparent 
dual superconductivity. The latter has  been recognized since long 
as an explanation of confinement~\cite{'tHooft:pu
} 
and is also intensively studied on the lattice~\cite{DiGiacomo:2002mm}. 
These investigations seem to indicate that the fundamental dual excitations are monopoles or vortices, 
while instantons might be composite objects of those.

The QCD vacuum structure may be studied on the lattice, starting from first principles. 
Alternatively, low-energy phenomenology
can be exploited to learn about the vacuum, at the expense of introducing 
model assumptions as for example 
in the instanton liquid 
model~\cite{Shuryak:1981ff
} or in the
model of the stochastic vacuum~\cite{Dosch:ha}, which incorporates the 
dual superconductor picture of the vacuum.

In this short review, I shall emphasize that manifestations of the vacuum structure
of non-Abelian gauge theories 
can also be searched for in high-energy scattering. 
I shall concentrate in Sect.~\ref{ihard} on instanton-induced hard scattering processes 
in the electroweak gauge theory 
(Quantum Flavor Dynamics (QFD))~\cite{Ringwald:1989ee,
Ringwald:1990qz,
Zakharov:1990dj,
Khoze:1991mx,Morris:1993wg,Rubakov:1996vz} and in deep-inelastic scattering  (DIS)  
in  QCD~\cite{Ringwald:1994kr,Moch:1996bs,Ringwald:1998ek,Ringwald:1999ze,Ringwald:1999jb,Ringwald:2000gt},  
which are calculable from first principles within instanton-perturbation theory. In Sect.~\ref{soft}, 
I shall discuss briefly soft scattering processes in QCD.

\section{\label{ihard}HARD SCATTERING PROCESSES INDUCED BY INSTANTONS}

Instantons ($I$) are minima of the Euclidean action, localized in space and Euclidean
time, with unit 
topological charge $Q=1$. In Minkowski space-time, they 
describe tunneling transitions between classically degenerate, topologically
inequivalent vacua, differing in their 
winding number
by one unit, $\triangle N_{\rm CS}=Q=1$~\cite{Jackiw:1976pf
}. 
The corresponding energy barrier, under which the instantons tunnel,  
is inversely proportional to the gauge coupling $\alpha_g$ and the effective $I$-size $\rho_{\rm eff}$, 
$M_{\rm sp}\sim \pi/(\alpha_g\,\rho_{\rm eff})$, and of order 
$\pi M_W/\alpha_W\sim 10$ TeV in QFD~\cite{Klinkhamer:1984di} and ${\mathcal Q}$ in hard scattering 
in QCD, where ${\mathcal Q}\gwig 10$ GeV is a large momentum transfer 
e.g. in DIS~\cite{Ringwald:1994kr,Moch:1996bs,Ringwald:1998ek}.

\subsection{Instanton-induced processes}

Due to axial anomalies~\cite{Adler:gk
}, $I$-induced hard scattering processes 
are always associated with anomalous fermion-number violation~\cite{'tHooft:fv}, in particular baryon plus 
lepton number 
violation, $\triangle ( B+L )=-12\, Q$, in the case of QFD, chirality violation,
$\triangle Q_5 = 2\,n_f\,Q$, in the case of QCD. 
Generically, $I$-induced total cross-sections for hard parton scattering processes
are given in terms of an integral over the instanton-antiinstanton ($I\bar I$) collective coordinates 
(sizes $\rho,\bar\rho$, $I\bar I$ distance $R$, color 
orientation $U$)~\cite{Ringwald:1998ek} (see also~\cite{Zakharov:1990dj,
Khoze:1991mx,Balitsky:1992vs})
\begin{eqnarray}
\nonumber
\lefteqn{
      \sigma^{(I)}_{p_1p_2} \sim }
\\ \nonumber 
&&
      \int d^4 R
      \int\limits_0^\infty d\rho 
      \int\limits_0^\infty d\overline{\rho}\, 
      { D(\rho) D(\overline{\rho})}
       \int dU
            {\rm e}^{{-\frac{4\pi}{\alpha_g}}
      { \Omega\left(U, \frac{R^2}{\rho\bar\rho},\frac{\bar\rho}{\rho} \right)}}
\\ \label{gencross}
&&
   {\rm e}^{{\rm i} ( p_1+p_2 )\cdot R
                -\sum_{i=1}^2 \sqrt{-p_i^2}\,(\rho +\bar\rho )} 
\,\ldots
\,. 
\end{eqnarray}
Here, the basic blocks arising in $I$-per\-tur\-bation theory 
are the $I$-size distribution $D(\rho )$
and the
function $\Omega$, which takes into account the exponentiation of gauge boson 
production~\cite{Ringwald:1989ee
} and can be identified with the 
$I\bar I$-interaction defined via the valley method~\cite{Yung:1987zp,Khoze:1991mx,Verbaarschot:1991sq}.

\begin{figure}[h]
\vspace{-5ex}
\begin{center}
 \includegraphics*[width=4.5cm,bbllx=0pt,bblly=0pt,bburx=318pt,bbury=232,clip=]{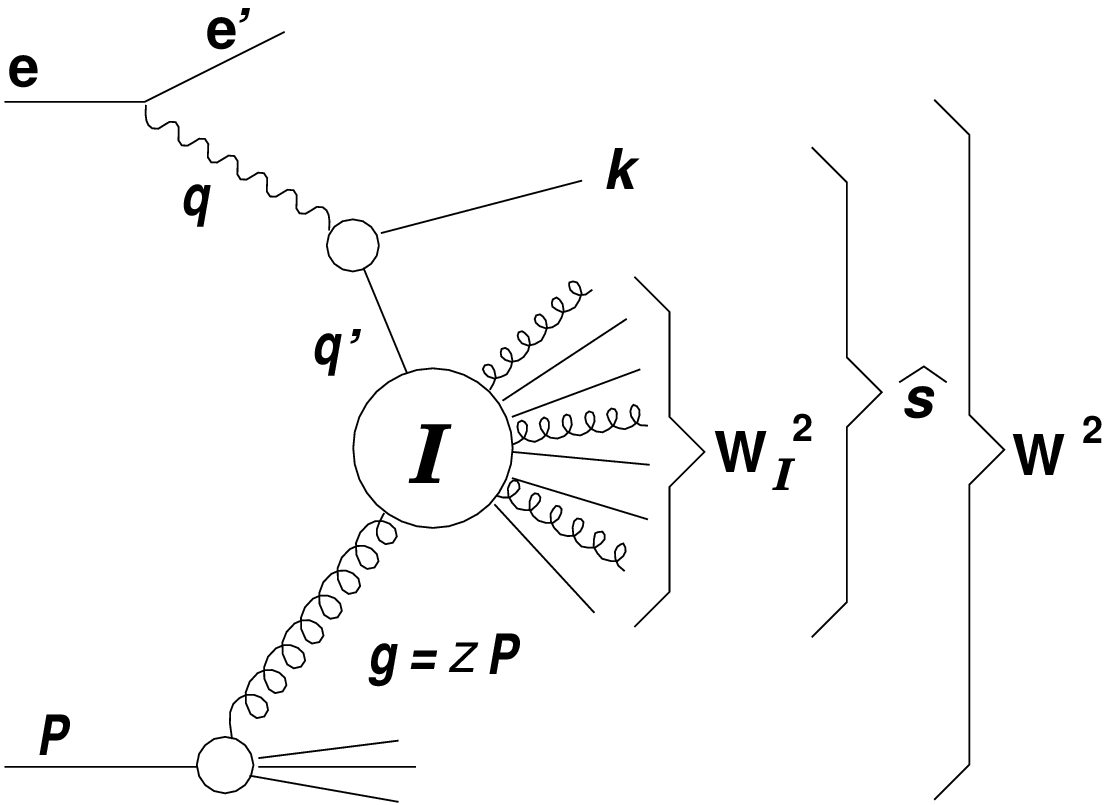}
\vspace{-5ex}
   \caption[dum]
     {
     QCD $I$-induced process in DIS.
   \label{kin-var}}
\end{center}
\end{figure}
\begin{figure} [h]
\vspace{-5.7ex}
\begin{center}
\parbox{3.6cm}{\includegraphics*[width=3.7cm]{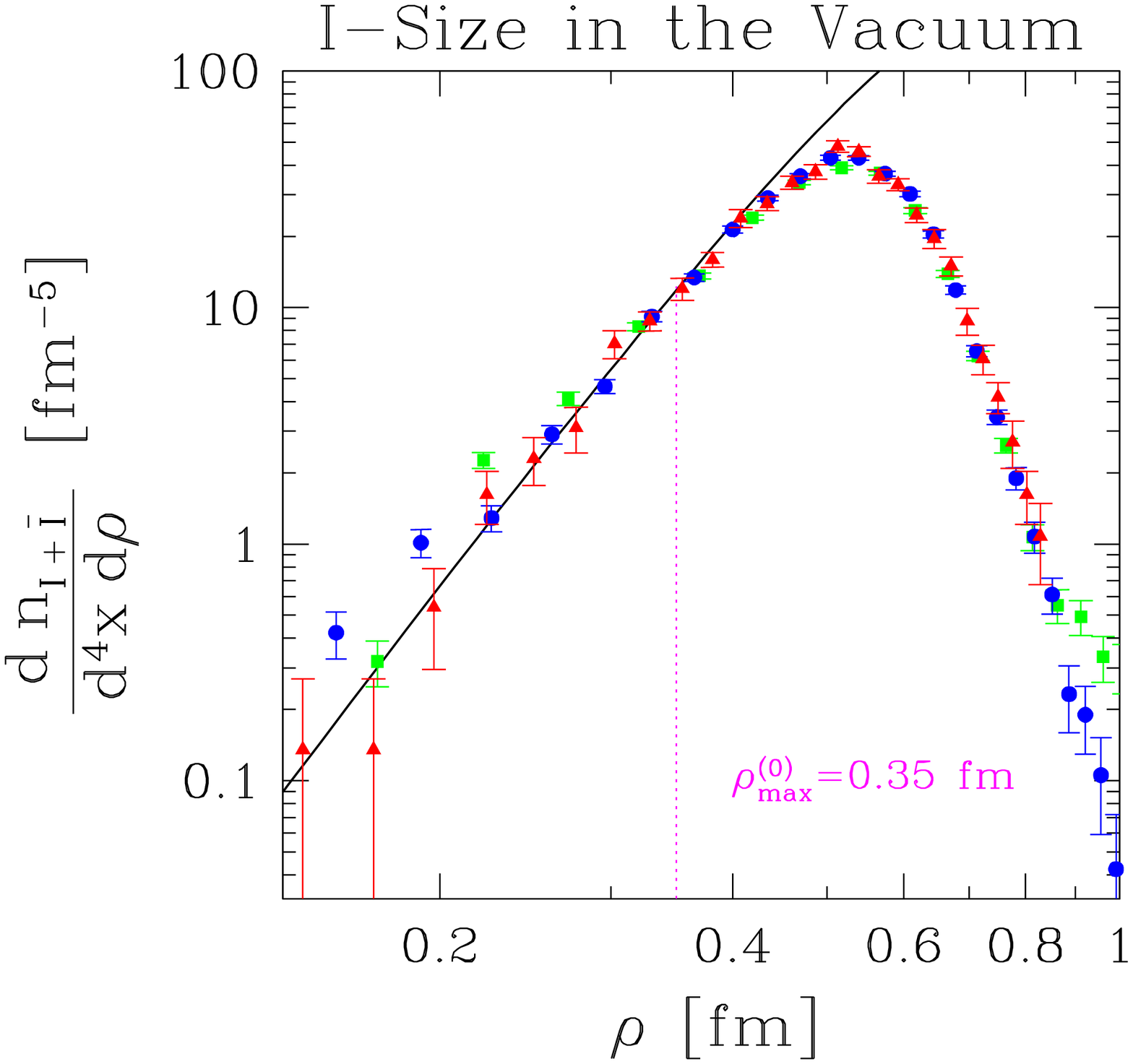}}
\parbox{3.6cm}{\includegraphics*[width=3.7cm]{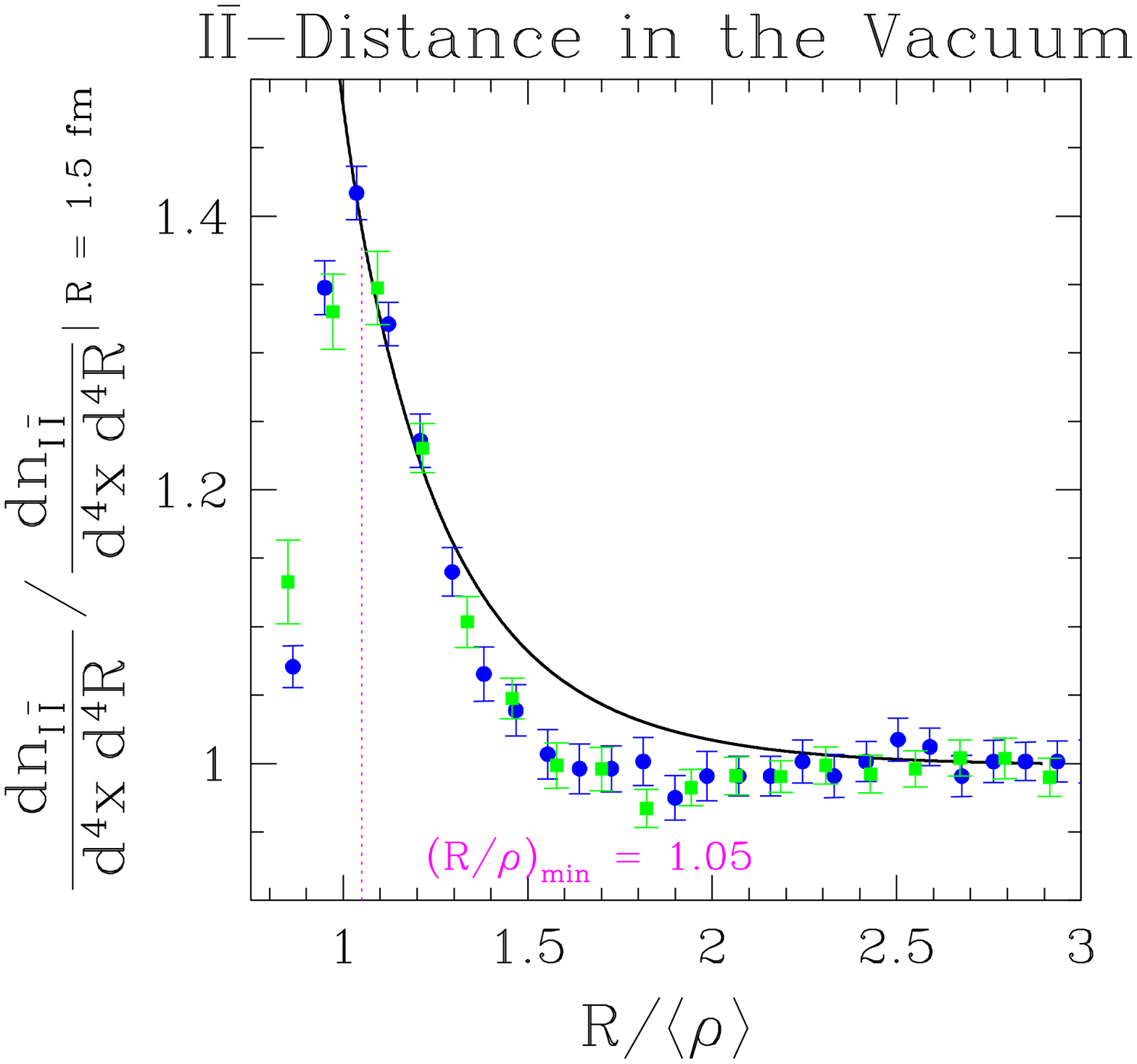}}\\[2ex]
\vspace{-7ex}
\caption[dum]{
Illustration of the agreement of 
lattice data~\cite{Smith:1998wt} 
for the $I$-size distribution (left) and the $I\overline{I}$-distance
distribution (right) with the predictions from  
$I$-perturbation theory for $\rho\lwig
0.35$ fm and $R/\rho\gwig 1.05$, respectively~\cite{Ringwald:1999ze,Ringwald:2000gt}. 
\label{lattice}}  
\end{center}
\end{figure}

\vspace{-8ex}  
The $I$-size distribution $D(\rho )$ is known in 
$I$-perturbation theory, $\alpha_g (\mu )\ln (\rho\mu )\ll 1$,
up to two-loop renormalization group invariance~\cite{'tHooft:fv,Bernard:1979qt
}. 
At one-loop, it reads  
     \begin{eqnarray}
\label{size-dist}
     { D({\rho})} =
     \frac{d}{\rho^5}
      \left(\frac{2\pi}{\alpha_g (\mu )}\right)^{2\,N_c} 
      (\rho\,
      \mu )^{\beta_0\,S^{(I)}
      } \ 
     {\rm e}^{-\frac{2\pi}{\alpha_g (\mu )}\,{ S^{(I)}}}     
      \end{eqnarray} 
with $\beta_0 = 11\,N_c/3 -2\,n_f/3 -1/6\,n_s$ the first coefficient in the 
$\beta$ function, 
\begin{eqnarray}
\label{action}
      { S^{(I)}} =
               \cases {1 &{\rm QCD}\,;\cr
                        1 + M_W^2\,\rho^2/2 & {\rm QFD}\,, \cr}
\end{eqnarray}
the $I$-action, $\mu$ the renormalization scale, and $d$ a scheme-dependent constant. 
The quite differenct $\rho$ dependence of 
the size distribution~(\ref{size-dist}) for QCD and QFD has important consequences 
for the predictivity: whereas large-size instantons, $\rho\gwig M_W^{-1}$, are exponentially 
suppressed in QFD (cf.~(\ref{action})) and thus the relevant contributions to the
size integrals in~(\ref{gencross}) arise consistently from the perturbative region 
($\alpha_W(\rho^{-1})\ll 1$) even if both initial partons are on-shell, $p_i^2\approx 0$, 
in QCD the power-law behavior of the size distribution,
$\sim \rho^{\beta_0 -5}$,
generically causes the dominant contributions to~(\ref{gencross}) to originate from 
large $\rho\sim \Lambda^{-1}\Rightarrow \alpha_s(\rho^{-1})\sim 1$ and thus
often spoils the applicability of $I$-perturbation theory.
In DIS  (cf.\,Fig.\,\ref{kin-var}), however, one parton, $p_1= q^\prime$ say, 
carries a space-like virtuality $Q^{\prime 2}=-p_1^2>0$, such that  
the contribution of large
instantons to the integrals is suppressed by an exponential
factor in~(\ref{gencross}), ${\rm e}^{-Q^\prime (\rho+\bar\rho )}$,    
and $I$-perturbation theory becomes exploitable\ifhepph, i.e. predictable,\fi  
\ for sufficiently large $Q^\prime$~\cite{Moch:1996bs,Ringwald:1998ek}. 
In this connection it is quite welcome that 
lattice data on the instanton content of the quenched ($n_f=0$) 
QCD vacuum~\cite{Smith:1998wt}
can be exploited to infer the region of validity of $I$-perturbation
theory for $D(\rho )$~\cite{Ringwald:1999ze,Ringwald:2000gt}: 
As illustrated in Fig.~\ref{lattice} (left), there is very good agreement for  $\rho\lwig 0.35$ fm. 

\vspace{-7.5ex}
As far as the $I\bar I$-interaction $\Omega$ is concerned, it is seen from
simple Fourier correspondences in~(\ref{gencross}), e.g. $R^2\sim 1/(p_1+p_2)^2\equiv 1/s^\prime$, 
that at high center-of-mass (cm) energies $\sqrt{s^\prime}$ small $I\bar I$-distances are 
probed, i.e. $\rho\bar\rho /R^2\sim s^\prime/Q^{\prime 2}\equiv 1/x^\prime -1$ (DIS) 
or $\sim s^\prime/M_W^2$ (QFD). 
Thus, in order to make a reliable prediction of $I$-induced hard scattering at high energies, 
we need to know the interaction for small distances. Again, for QCD one may exploit 
lattice data on the $I\bar I$-distance distribution~\cite{Smith:1998wt} to infer the region of validity
of the description of the $I\bar I$-interaction by its exact expression given in the valley 
method~\cite{Khoze:1991mx,Verbaarschot:1991sq}: one finds good agreement for 
$R/\rho\gwig 1.0\div 1.05$~\cite{Ringwald:1999ze,Ringwald:2000gt} (cf. Fig.~\ref{lattice}). 
In this case, however, there are remaining theoretical ambiguities: The integrations over $\rho$, $\bar\rho$ 
in the $I\bar I$-distance distribution imply
significant contributions also from larger instantons with 
$0.35 {\rm \ fm}\lwig\rho,\overline{\rho}\lwig 0.6$ fm, outside the strict region of 
perturbation theory. It is not excluded that the valley interaction reliably describes the interactions
of small-size instantons, $\rho\sim\bar\rho\ll\langle\rho\rangle \approx  0.5$ fm, at smaller
$R/\rho$, say~$R/\rho\gwig 0.5$.    

\begin{figure} [h] 
\vspace{-6ex} 
\begin{center}
\includegraphics*[angle=-90,width=5cm]{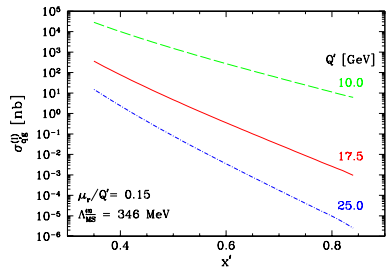}
\vspace{-5ex}
  \caption[dum]{QCD $I$-subprocess cross-section~\cite{Ringwald:1998ek,Ringwald:1999jb}.  
\label{isorho}}
\end{center}
\end{figure}
\begin{figure} [h]
\vspace{-14ex}
\begin{center}
\includegraphics*[width=6cm,clip=]{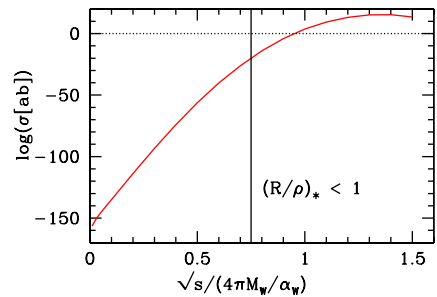}
\vspace{-12ex}
\caption[dum]{QFD $I$-subprocess cross-section~\cite{Khoze:1991mx,Ringwald:2002qfd}.
\label{vlhc}}  
\end{center}
\end{figure}

\vspace{-10ex}
The energy dependencies of the $I$-induced parton-parton cross sections illustrated 
in Fig.~\ref{isorho} for DIS ($x^\prime = Q^{\prime 2}/(Q^{\prime 2}+s^\prime )$)~\cite{Ringwald:1998ek,Ringwald:1999jb}  
and in Fig.~\ref{vlhc}  for QFD~\cite{Khoze:1991mx,Ringwald:2002qfd}    
are easily understood on the basis of above mentioned Fourier correspondences and associated 
saddle-point relations: At increasing energies, smaller and smaller  $I\bar I$-distances are probed, and 
the cross-sections are rapidly growing due to the attractive nature of the $I\bar I$-interac\-tion  in the 
perturbative semi-classical regime (cf. Fig.~\ref{lattice} (right)). 
This goes along with the production of $n_g\sim 1/\alpha_g$ gauge bosons, 
in addition to the minimal number of fermions required by the axial
anomaly.
  Furthermore,  in the case of DIS, 
at increasing virtuality $Q^{\prime}$ 
smaller and smaller instantons are probed and the cross-section diminishes  
in accord with the size distribution (Fig.\,\ref{lattice}~(left)).

\subsection{QCD-instantons at HERA}

With my colleague Fridger Schrempp we have conducted a long-term research program 
at DE\-SY to work out the theory and phenomenology of hard 
QCD $I$-induced processes in DIS at 
HERA~\cite{Ringwald:1994kr,Moch:1996bs,Ringwald:1998ek,Ringwald:1999ze,Ringwald:1999jb,Ringwald:2000gt}.  
Meanwhile, the first dedicated search by the H1 collaboration has been 
published~\cite{Adloff:2002ph}.  
Several observables characterising the hadronic final state of QCD $I$-induced
events were exploited to identify a potentially $I$-enriched domain. The results obtained
are intriguing but non-conclusive. While an excess of events with $I$-like topology
over the expectation of the standard DIS background is observed,  which, moreover,  is compatible 
with the $I$-signal,   it can not be claimed to be significant
given the uncertainty of the Monte Carlo simulations of the standard DIS background.
The data do not exclude the cross-section predicted by $I$-perturbation theory 
for quite small $I$-sizes $\rho_\ast\lwig 0.2$ fm and $(R/\rho )_\ast \gwig 0.9$. 

\vspace{-12ex}
\subsection{QFD-instantons at VLHC?}

In the early `90s the possibility of observable QFD $I$-effects at cm energies $\gg 10$ 
TeV was quite intensively investigated~\cite{Ringwald:1989ee,
Ringwald:1990qz,
Zakharov:1990dj,
Khoze:1991mx,Morris:1993wg,Rubakov:1996vz}. But despite considerable theoretical efforts the 
actual size of the cross-sections in the relevant energy regime was never established. In view of the similarity
between QFD and hard QCD $I$-induced processes in DIS and of the recent
information about the latter both from the lattice and from experiment, it
seems worthwhile to reconsider the subject~\cite{Ringwald:2002qfd}.    
Figure~\ref{vlhc} represents a state-of-the-art evaluation of the QFD $I$-induced 
quark-quark cross-section, which neglects Higgs production~\cite{Khoze:1991mx}. 
It demonstrates that the cross-section is unobservably small in the conservative
fiducial region corresponding to $(R/\rho )_\ast >1$. However, it becomes of observable 
size if the $I\bar I$-attraction remains valid also for slightly smaller values of 
$(R/\rho )_\ast\gwig 0.7$ at $\sqrt{s^\prime}\approx 4\pi M_W/\alpha_W = 30$ TeV.  
This opens up exciting opportunities at future colliders~\cite{Ringwald:1990qz
} such as VLHC~\cite{vlhc}
or at cosmic ray facilities and neutrino telescopes~\cite{Morris:1993wg}. 

\section{\label{soft}SOFT HIGH-ENERGY PROCESSES}

One of the main challenges of QCD is to understand the high-energy, but small momentum transfer
cross-sections of hadrons from first principles. In this domain, perturbation theory is not 
applicable  and direct lattice simulations are not possible until now.
There have been recently several attempts to face this challenge~\cite{Hebecker:2001ex}. 
Most of them are based on the eikonal representation of the amplitudes in terms of correlators
of Wilson lines or loops~\cite{Nachtmann:ua}, which thereafter are evaluated using different 
model assumptions. The most advanced attempt is certainly the one of the Heidelberg 
group~\cite{Dosch:ym
} based on the model of the stochastic vacuum~\cite{Dosch:ha}, which has been already successfully applied to a variety of
soft high-energy processes (e.g.~\cite{Dosch:2000jg
}).    
Less developed but rapidly evolving is the understanding of a possible connection of 
semi-hard instantons to diffraction~\cite{Kharzeev:2000ef
}.  There exists the intriguing possibility that larger-size instantons 
build up diffractive scattering, with the marked $I$-size scale $\langle \rho\rangle \approx 0.5$ fm 
(Fig.~\ref{lattice} (left)) being reflected in the conspicuous ``geometrization'' of soft 
QCD~\cite{Schrempp:2002kd}.

\end{document}